\documentclass[twocolumn,prd,nofootinbib,aps,prl,floats,floatfix,amsmath,amssymb,longbibliography,secnumarabic]{revtex4-1} %
\usepackage[english]{babel}

\usepackage[letterpaper,top=2cm,bottom=2cm,left=3cm,right=3cm,marginparwidth=1.75cm]{geometry}

\usepackage{amsmath}
\usepackage{graphicx}
\usepackage[colorlinks=true,citecolor=red]{hyperref}

\newcommand{\be}{\begin{equation}}
\newcommand{\ee}{\end{equation}}

\newcommand{\bea}{\begin{eqnarray}}
\newcommand{\eea}{\end{eqnarray}}

\begin{document}

\title{Comment on ``Dynamical Dark Energy at Late Time $\Lambda$CDM''}
\author{James M.\ Cline}
\email{jcline@physics.mcgill.ca}
\affiliation{McGill University Department of Physics \& Trottier Space Institute, 3600 Rue University, Montr\'eal, QC, H3A 2T8, Canada}

\begin{abstract}
In \url{https://arxiv.org/abs/2505.18900} it was claimed that an apparent evolution of the dark energy equation of state occurs within the standard
$\Lambda$CDM cosmological model.  I point out that this erroneous conclusion is due to a mathematical error.

\end{abstract}

\maketitle

Recently Ref.\ \cite{Moffat:2025jmx} derived a relation between the dark energy equation of state $w(z)$ and the Hubble parameter $H(z)$ which purported to show that even if the dark energy is a cosmological constant, $w(z)$ will vary with redshift.  This conclusion is quite surprising, so it comes as no surprise to discover that it is false, arising from making inconsistent approximations.

The authors start by almost writing a correct relation,\footnote{they wrote a dimensionally incorrect equation, which appears to be due to misplacing a brace in LaTeX.  The corrected version is given here.} valid at late times when radiation can be ignored,
\be
    w(t) = -{2\dot H + 3 H^2\over 3 H^2 -8\pi G\rho_m}\,,
\ee
where $\rho_m$ is the matter density.
They then proceed to neglect $\rho_m$, presuming that the Universe is dark energy dominated at late times.  This is the mistake.  Trading $t$ for redshift, they obtain the wrong result
\be
    w(z) = -1 + {2(1+z)\over 3 H}\,{dH\over dz}
\ee
instead of the correct one
\be
    w(z) = {2(1+z) H {dH\over dz} - 3 H^2\over 3 H^2-3\pi G\rho_m} = -1\,.
\ee

\bibliographystyle{utphys}
\bibliography{sample}

\end{document}